\documentclass[aps,prl,twocolumn,floatfix,superscriptaddress,showpacs,amssymb,amsmath]{revtex4}
\usepackage{graphicx}

\newcommand*{\AMeV}{\textit{A}\;MeV}

\begin{document}



\title{ Distinctive features of Coulomb-related emissions
        in peripheral heavy ion collisions at Fermi energies}


%
%

\author{S.~Piantelli},
\affiliation{Sezione INFN di Firenze, 
             Via G. Sansone 1, I-50019 Sesto Fiorentino, Italy}

\author{P.R.~Maurenzig},
\affiliation{Sezione INFN and Universit\`a di Firenze, 
             Via G. Sansone 1, I-50019 Sesto Fiorentino, Italy}

\author{A.~Olmi},
\thanks{corresponding author}
\email[e-mail:]{olmi@fi.infn.it}
\affiliation{Sezione INFN di Firenze, 
             Via G. Sansone 1, I-50019 Sesto Fiorentino, Italy}

\author{L.~Bardelli},
\affiliation{Sezione INFN and Universit\`a di Firenze, 
             Via G. Sansone 1, I-50019 Sesto Fiorentino, Italy}

\author{M.~Bini},
\affiliation{Sezione INFN and Universit\`a di Firenze, 
             Via G. Sansone 1, I-50019 Sesto Fiorentino, Italy}

\author{G.~Casini},
\affiliation{Sezione INFN di Firenze, 
             Via G. Sansone 1, I-50019 Sesto Fiorentino, Italy}

\author{A.~Mangiarotti},
\thanks{present address:
        Laborat\'orio de Instrumenta\c{c}\~{a}o e 
        F\'{\i}sica Experimental de Part\'{\i}culas, 3004-516 
       Coimbra, Portugal}

\affiliation{Sezione INFN and Universit\`a di Firenze,
             Via G. Sansone 1, I-50019 Sesto Fiorentino, Italy}

\author{G.~Pasquali},
\affiliation{Sezione INFN and Universit\`a di Firenze, 
             Via G. Sansone 1, I-50019 Sesto Fiorentino, Italy}

\author{G.~Poggi},
\affiliation{Sezione INFN and Universit\`a di Firenze, 
             Via G. Sansone 1, I-50019 Sesto Fiorentino, Italy}

\author{A.A.~Stefanini},
\affiliation{Sezione INFN and Universit\`a di Firenze, 
             Via G. Sansone 1, I-50019 Sesto Fiorentino, Italy}

\date{\today}

\begin{abstract}
Light charged particles emitted at about $90^{\circ}$ in the frame of 
the projectile-like fragment in semi-peripheral collisions of 
$^{93}$Nb+$^{93}$Nb at 38\AMeV\ give evidence for the simultaneous 
occurrence of two different production mechanisms.
This is demonstrated by differences in the kinetic energy spectra and 
in the isotopic composition of the particles.
The emission with a softer kinetic energy spectrum and a low N/Z ratio 
for the hydrogen isotopes is attributed to an evaporation process. 
The harder emission, with a much higher N/Z ratio, 
can be attributed to a ``midvelocity'' process consisting of a 
non-isotropic emission, on a short time-scale, 
from the surface of the projectile-like fragment.
\end{abstract}


\maketitle

In peripheral and semiperipheral heavy ion collisions at Fermi energies 
a sizable fraction of the emitted light charged particles (LCPs) and 
intermediate mass fragments (IMFs, $Z\geq$3) are produced at 
``midvelocity'', i.e., they have velocities intermediate between 
those of the projectile-like fragment (PLF) and of the target-like fragment 
(TLF) (see, e.g., 
\cite{Lukasik97,Plagnol99,Piantelli02,Mangiarotti04,Piantelli06,Milazzo02,Defilippo05neck} 
and references therein).
The midvelocity emissions represent a new, distinctive feature of the 
Fermi energy domain.
However, in spite of many efforts, a comprehensive description of its 
characteristics and evolution with bombarding energy is still lacking, 
and also the nature of its production mechanism, whether statistical or 
dynamical, is still a debated matter.

One of the main difficulties, especially on the low energy side of the Fermi
energy region, is separating the midvelocity emissions from the 
evaporation of PLF and TLF.
For this purpose, one usually assumes a nearly isotropic distribution 
of the evaporated LCPs in the rest frame of the emitting PLF (and TLF)
(indeed, for moderate spins, the out-of-plane anisotropy is rather weak).
Then, the procedure (see, e.g., \cite{Plagnol99,Piantelli06}) 
consists in 
attributing the forward emissions in the PLF frame to an isotropic evaporation 
from the PLF, extrapolating it to all angles and finally subtracting it 
from the total measured distribution to obtain the midvelocity component.
The resulting emission pattern for midvelocity products is usually
displayed in the ($v_{\,\parallel}, v_{\,\perp}$) plane, where 
$v_{\,\parallel}$  ($v_{\,\perp}$) is the parallel (perpendicular) component 
of their c.m. velocity with respect to 
the PLF-TLF separation axis (or, sometimes, the beam axis).

Close examination of this emission pattern 
suggests that the midvelocity emissions consist
not only of a broad distribution roughly centered at velocities intermediate
between PLF and TLF, but also of an anisotropic 
``surface'' emission \cite{Piantelli02,Piantelli06}
located along the Coulomb ridge of the PLF (and of the TLF as well).
One can imagine different mechanisms which might contribute.
Early emissions --- like, e.g. 
pre-equilibrium particles from the very first phases of the collision, or
particles from the hot zone of overlapping matter during contact ---
are expected to populate mainly the central midvelocity region, possibly 
with large transverse momenta.
Later emissions in the separation phase (or immediately after separation)
are likely to display an increasingly strong relationship with just one of the 
reaction partners, so they are expected to show a kind of partial ``orbiting'' 
and to be mainly distributed, in a non-isotropic way, on the Coulomb ridge of 
one reaction partner (usually experiments concentrate on the PLF), 
preferentially facing the other one (the TLF).
These later emitted Coulomb-related particles may just be neck remnants, 
left behind by a dynamical multiple neck-rupture process \cite{Colin03}; 
or they may be produced, after neck rupture, by the fast non-equilibrated 
decay of a possibly strongly deformed PLF, in a process 
resembling a fast oriented fission \cite{Stefanini95,Bocage00,Defilippo05ff} 
for extreme mass asymmetries \cite{Piantelli02,Defilippo05neck}.
Opposite interpretations ascribe these later emissions to a pure statistical 
evaporation from the PLF, however perturbed by the proximity of the TLF
(\cite{Hudan04,Jandel05}).   

In this Letter we put into evidence some distinctive features of 
this ``surface'' midvelocity component, which has been measured in a common 
angular range simultaneously with the PLF evaporation,
in particular for what concerns the kinetic energy spectra of  
LCPs and the average isotopic composition of the $Z=1$ particles.  
They may help constraining models on midvelocity processes.

The experimental data refer to the symmetric collision 
$^{93}$Nb+$^{93}$Nb at 38\AMeV, studied with the \textsc{Fiasco} setup
(for details see \cite{Bini03,Piantelli02,Piantelli06})
at the Superconducting Cyclotron of the Laboratori Nazionali 
del Sud of INFN in Catania.
Attention is focused on two-body semi-peripheral reactions,
where the PLF and TLF emission patterns can be clearly separated.
These reactions are selected by requiring only two heavy 
($Z\gtrsim10$) fragments in the exit channel, namely the PLF and TLF 
(the latter one is efficiently detected even in the most peripheral 
collisions due to the low thresholds of the gas detectors).

For sorting purposes, we use the ``Total Kinetic Energy Loss'', 
a parameter obtained from the kinematic coincidence analysis \cite{Casini89} 
and defined as  
$\mathrm{TKEL}  = \mathrm{E}_\mathrm{\,in}^\mathrm{\;c.m.}\, -\,
   \tfrac{1}{2}  \,  \Tilde{\mu}\, { v_{\mathrm{rel}}^{\,2}}$,
where $E_\mathrm{\,in}^\mathrm{\;c.m.}$ is the center-of-mass
energy in the entrance channel,  
$v_{\mathrm{rel}}$ the PLF-TLF relative velocity and 
$\Tilde{\mu}$ the reduced mass assuming an exactly binary 
reaction.
As noted in \cite{Piantelli06}, at Fermi energies TKEL does not 
represent any more a good estimate of the true total kinetic energy 
loss of the collision, but it is used just as an
\textit{ordering parameter for sorting the events in bins of increasing
centrality}.  
More details on the use of TKEL as an impact parameter estimator 
can be found in Appendix A of \cite{Piantelli06}. 

As the solid angle coverage of \textsc{Fiasco}, although large,  
is significantly smaller than 4$\pi$, the yields of LCPs and IMFs are 
first corrected \cite{Piantelli06} for the limited geometrical coverage
and for the low-energy identification thresholds.
Then, using relativistic kinematics
 \footnote{
A relativistic transformation, instead of a Galileian one, is 
  preferred to avoid distortions of the angular distribution of the fastest 
  particles, in particular for protons, 
  which may reach lab-velocities as large as $\beta$ = 0.3.},
we obtain the emission patterns in the PLF frame 
(the polar axis is the asymptotic PLF-TLF separation axis
pointing in the direction of motion of the PLF).
One can now examine the distribution of the parallel and perpendicular 
components of the reduced particle momenta with respect to the polar axis 
($\frac{p_{\,\parallel}}{m}$ and $\frac{p_{\,\perp}}{m}$, respectively). 
As shown in Fig. 4 of \cite{Piantelli06}, the most forward 
part of the emission can be ascribed mainly to an evaporative process. 
On the contrary, at backward angles, the yields of all particle species 
present a large excess, with respect to a pure evaporative process,
which is globally ascribed to midvelocity emissions, with a tail of
the ``surface'' component actually extending at polar angles
well below $\theta=90^\circ$.

\begin{figure}[t]
  \includegraphics[width=73mm,bb= 0 0 530 485,clip]{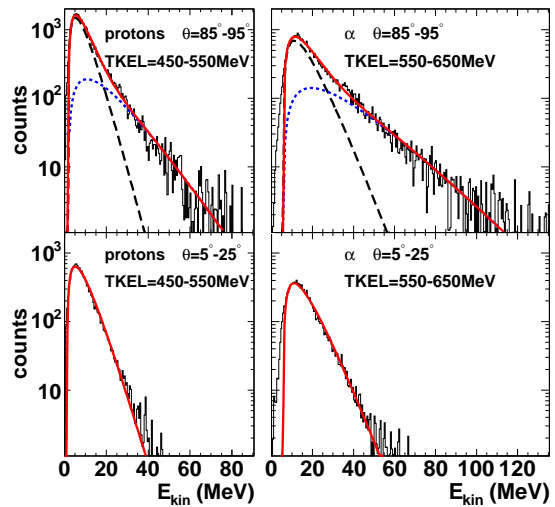}
  \caption
  {\label{fig1}
  (color online)
  Kinetic energy spectra, in the reference frame of 
  the PLF, for protons (left panels) at TKEL = 450--550 MeV 
  and $\alpha$ particles (right panels) at 550--650 MeV.
  Upper and lower panels refer to particles emitted in the 
  angular range $85^{\circ}\leq \theta \leq 95^{\circ}$ and
  $5^{\circ}\leq \theta \leq 25^{\circ}$, respectively.
  The full (red) curves are fits to the data
  with one effusive Maxwell function (lower panels) and with
  the superposition of two effusive Maxwell functions (upper panels), 
  indicated by the dashed (black) and dotted (blue) curves.
  }
 \end{figure}

The kinetic energy distributions of the particles in the PLF frame
show clear distinctive features, which are best observed around 
$90^{\circ}$ where the contribution of the early central emissions is 
strongly reduced, an appreciable contribution of ``surface'' midvelocity 
particles is still present and the emissions from the TLF are negligible.
For example, the upper panels of Fig. \ref{fig1} show the spectra of protons 
at TKEL = 450--550 MeV (left) and $\alpha$ particles at 550--650 MeV (right).
For comparison, the lower panels show the corresponding spectra for  
forward emitted particles.
All spectra have been corrected for the angle-dependent effects
caused by the recoil of the emitter 
(see \cite{Piantelli06} for more details).
Such recoil effects are however relatively small at forward angles 
and become negligible around $\theta=90^{\circ}$.

The distributions of forward emitted particles of the lower panels in 
Fig. \ref{fig1} have been fitted with a single effusive Maxwell function 
(red curve)
\begin{equation}
  f_0=N_0\: [(E-Z B_0)/T_0^2] \; \exp[-(E-Z B_0)/T_0],
\label{eq1}
\end{equation}
with $Z$ charge of the emitted particle, $B_0$ Coulomb barrier divided 
by $Z$ and $T_0$ (inverse) slope parameter of the distribution.
On the contrary, the distributions in the upper panels
cannot be fitted with a single Maxwell function, but they need a 
two-component function 
\begin{eqnarray}
 f_{12}&=& N_1\: [(E-Z B_1)/T_1^2] 
              \: \exp[-(E-Z B_1)/T_1] \nonumber \\
       &+& N_2\: [(E-Z B_2)/T_2^2] 
              \; \exp[-(E-Z B_2)/T_2] 
\label{eq2}
\end{eqnarray}
with two different slope parameters $T_1$ and $T_2$ ($T_1 < T_2$)
and equal barriers \footnote{
   The fit determines $B_2$ with large uncertainty and there are arguments 
   to expect both a higher (larger $Z$ of the emitter) and a 
   lower (deformed emitter) value than $B_1$, so we finally use $B_2=B_1$.}.
These two components are shown in Fig. \ref{fig1} by the dashed (black) and
dotted (blue) curves, respectively (the red full curve is their sum).
No attempts were made to reproduce the smearing
of the barriers, so the low-energy part of the histograms (below 
$\approx$ 70\% of the peak height) is not used in the fit. 
Fits of similar quality are obtained for all LCPs at all TKEL values.

\begin{figure}[t]
  \includegraphics[width=76mm,bb= 0 0 520 485,clip]{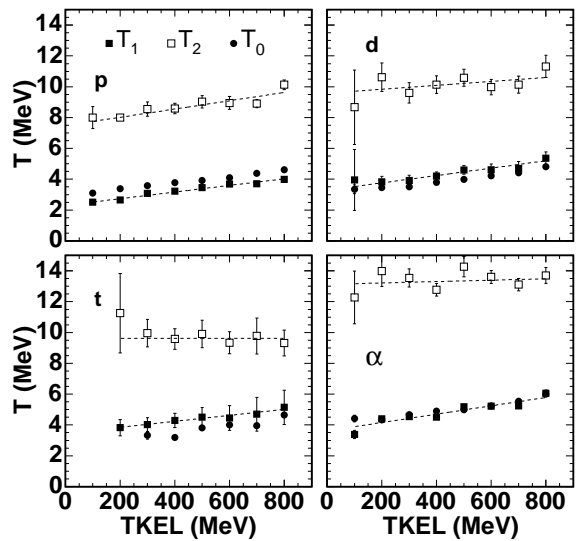}
  \caption
  {\label{fig2}
   Values of the slope parameters $T_0$ (filled circles) 
   $T_1$ (filled squares) and $T_2$ (open squares) as a function of 
   TKEL for protons, deuterons, tritons and $\alpha$ particles. 
   The bars indicate just the statistical errors. 
   Dashed lines through the filled and open squares are just to guide the eye.
  }
 \end{figure}

By simple inspection one sees that the slopes of the two components are 
quite different.
For a quantitative comparison Fig.~\ref{fig2} presents $T_1$, 
$T_2$ and $T_0$ for all LCPs as a function of TKEL.
The error bars are only statistical;
further uncertainties (due, e.g., to efficiency and recoil corrections
and to the choice of the fit region) are around
0.1--0.2 MeV for $T_0$ and 0.5--1.0 MeV for $T_1$ and $T_2$.

The values of the slope parameters of the harder component, $T_2$, 
are indeed definitely larger than $T_1$ and $T_0$.
They are similar for all $Z=1$ isotopes (around 8-11 MeV), while a still 
higher value (about 13 MeV) is found for $\alpha$ particles and also for IMFs
(not shown in the figure; in this latter case, a fit with a single $T_2$
component is more appropriate, as the evaporation of  
IMFs from the PLF is negligible).
These values compare well with those of \cite{Lefort00}, which however 
refer to the central part of the midvelocity emissions in a lighter system 
at higher bombarding energies.

The second item worth noting is the quite similar values of $T_0$ and $T_1$.
This leads to the conclusion that the softer component in the spectra 
at $\approx 90^{\circ}$ (slope $T_1$, black dashed curves in the upper
panels of Fig.~\ref{fig1}) can be attributed to the usual 
evaporation from the excited PLF.  
Actually, in the tail of the spectra at forward angles (lower panels of 
Fig.~\ref{fig1}) there is a small excess of energetic particles 
(for protons and $\alpha$ particles, but not for deuterons and tritons), 
which might be due to a small pre-equilibrium contribution.
A two-component fit with Eq.~(\ref{eq2}) would give a slope parameter of the 
evaporative component slightly smaller than the $T_0$ obtained with 
Eq.~(\ref{eq1}) and hence in even better agreement with $T_1$,
at the expense of an increased uncertainty in the parameters of the 
second component.

The slope parameters of the evaporative process, $T_0$ and $T_1$, 
display a weak, almost linear, increasing trend with increasing TKEL.
For all particles, the values are rather similar
and vary in the range from 3 to 6 MeV.
These values and their dependence on the violence of the 
collision are in good agreement with those of other studies 
focused just on the decay of projectile residues in peripheral 
collisions at Fermi energies \cite{Ma97,Steckmeyer01}.
In the first bins of TKEL, the values of $T_1$ are somewhat 
larger than those expected on the base of a simple 
Fermi-gas formula, using the estimate of $E^\star$ from \cite{Piantelli06}.
However, $T_1$ and $T_2$ are just slope parameters.
A connection with nuclear ``temperatures'' is particularly difficult
in peripheral collisions at Fermi energies, due to the finite width of 
the distributions both in initial mass and excitation energy of the 
emitters and to uncertainties in the 
determination of the origin of the PLF frame.
These effects are discussed in detail in Ref. \cite{Vient06}, 
with regard to the decay of PLFs produced in peripheral collisions  
at Fermi energies and detected with the INDRA multidetector.  

What we want to stress is the remarkable difference between the 
slopes $T_1$ and $T_2$, which is found for all LCPs irrespective of 
the value (and hence of the selectivity) of the adopted ordering parameter. 
This fact indicates that the physical mechanisms associated with the two
components always maintain different characteristics.
Going toward smaller values of the angle $\theta$ in the PLF frame, 
$T_1$ does not change, while $T_2$ shows a weak 
tendency to decrease, but it remains definitely higher than $T_1$.
At the same time, the percentage of Coulomb-related midvelocity 
emission decreases.
Around $90^\circ$ it amounts to about 25, 50, 60 and 30\% 
of the total emissions of p, d, t and $\alpha$, respectively, 
and it decreases to less than 10\% in all cases for $\theta<40^{\circ}$.

\begin{figure}[t]
  \includegraphics[width=80mm,bb= 0 0 515 345,clip]{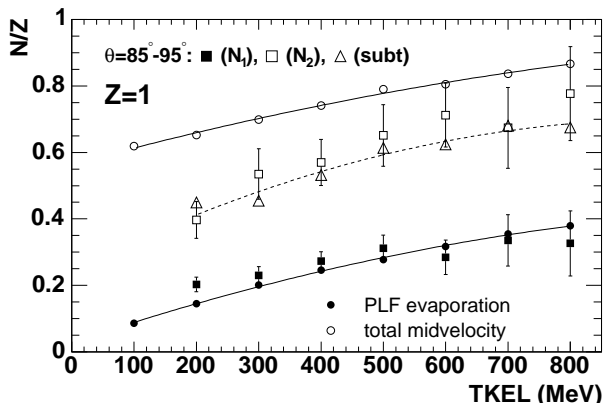}
  \caption
  {\label{fig3}
    N/Z ratio for hydrogens as a function of TKEL. 
    Open and full squares refer to the midvelocity and evaporative 
    components, respectively, and are obtained from fits of the kinetic 
    energy spectra at $85^{\circ} \leq \theta \leq 95^{\circ}$ 
    in the PLF frame.
    Open triangles refer to the midvelocity component
    in the same angular range and  
    are obtained by subtracting the evaporative component
    (see text); the dashed line is a quadratic fit to guide the eye.
    Full and open circles (with full lines) show 
    the results of \protect{\cite{Piantelli06}} for evaporation 
    and total midvelocity, respectively.} 
 \end{figure}

If the two components observed in the kinetic energy spectra correspond 
indeed to different production mechanisms, one may expect them to differ
also in other properties, like in the average 
isotopic composition of the emitted particles.
For this purpose we have estimated the N/Z ratio of the hydrogen isotopes
(for which the detectors have good isotopic resolution) in the angular 
range $85^{\circ}\leq\theta\leq 95^{\circ}$ in the PLF frame.
One can proceed in two different ways. 
The first one is similar to that adopted in Ref. \cite{Piantelli06}.
One first builds the whole evaporative invariant cross section,
$\sigma_{\mathrm{inv}}^{\mathrm{evap}} (\frac{p_{\parallel}}{m},\frac{p_{\perp}}{m})$, 
by rotating the part measured in the forward direction
so as to fill the whole angular range around the PLF position.
Then, one subtracts the so determined $\sigma_{\mathrm{inv}}^{\mathrm{evap}}$ 
from the measured total invariant cross section of particles emitted 
in the forward c.m. hemisphere, $\sigma_{\mathrm{inv}}^{\mathrm{tot}}$, 
thus finally obtaining an estimate of the yields of
midvelocity emissions of all types.
The results are shown by the open triangles in Fig.~\ref{fig3}.
The second completely different way to proceed is to deduce the average 
N/Z of the midvelocity emissions directly from the fitted intensity
parameter $N_2$: the results are shown by the open squares in Fig. \ref{fig3}.
This second method gives results in agreement with the previous one, 
although with larger uncertainties.
The other fit parameter, $N_1$, is used for an estimate of the isotopic 
composition of the evaporative component, shown by the filled squares.
Finally, in the same figure, the filled and open circles show the N/Z values 
obtained in Ref. \cite{Piantelli06} for the evaporative and total midvelocity 
emissions, respectively.

The average N/Z for the soft component of the kinetic energy 
spectra (filled squares) nicely agrees with the value obtained for the 
evaporation at forward angles (filled circles), thus landing further 
support to the identification of the soft component 
(with slope parameter $T_1$) with an evaporative process.
As shown in \cite{Piantelli06}, the experimental values
are indeed in close agreement with calculations employing
the statistical code \textsc{Gemini} \cite{Charity88b}.

For the Coulomb-related part of the midvelocity emissions, 
the open triangles are in a position intermediate between the 
open and filled circles at all TKEL.
This indicates that this Coulomb-related part of the midvelocity emissions,
although not as neutron rich as the bulk of the midvelocity emissions from 
the central ``source'', nevertheless presents an average isotopic composition
definitely higher than that of the usual evaporation 
(at least for the Z=1 particles).
This fact, while confirming a substantial difference between the
two mechanisms, may be taken as an indication of a possible
enrichment in bound neutrons of ``midvelocity'' matter 
\cite{Dempsey96,Lukasik97,Plagnol99,Larochelle00,Milazzo02,Piantelli02},
and/or it might be related to the reduced size of this ``source''
and hence to its higher energy concentration \cite{Mangiarotti04}.

In conclusion, we have investigated some aspects of LCPs 
emitted in peripheral collisions at Fermi energies. 
We have shown that, still at $90^{\circ}$ in the PLF frame, 
these emissions (with a clear origin from the PLF ``source'', as 
manifested by their concentration on the Coulomb 
ridge \cite{Piantelli02,Piantelli06}) consist of two components.
The softer component can be identified with the usual evaporation 
from an equilibrated source, 
on the basis of both the particle kinetic energies and 
the N/Z ratio of the hydrogen isotopes.  
The other component, which displays harder kinetic energy spectra of the 
particles, shows a more exotic N/Z, with a clear
neutron enrichment for bound neutrons with respect to evaporation.
These results represent a benchmark against which models describing 
midvelocity processes should be tested.
In any case the presented results are compatible with a picture in which 
the midvelocity ``surface'' emission comes from the highly excited contact 
region between PLF and TLF. 
After the neck rupture, the PLF may remain partially deformed and locally 
highly excited; as a consequence, it may tend to evaporate particles and 
fragments with high kinetic energy and with an isotopic composition 
resembling that of the neck region, before reaching 
full equilibration.  



%
 \bibliography{short}
%

\end{document}